\documentclass[11pt]{article}
\usepackage{epsfig}

\begin{document}

\baselineskip 30pt

\begin{center}
\begin{LARGE}
{\bf Photoemission From A Two Electron Quantum Dot
                      }
\end{LARGE}
\end{center}

\vskip 0.2in

\begin{large}
\begin{center}
Subinoy Das  \\
Department of Physics, Indian Institute of Technology, Kanpur, U.P.--208016, India \\
email: subinoy@iitk.ac.in   \\ \mbox{} \\
Pallab Goswami \\
Department of Physics, Indian Institute of Technology, Kanpur, U.P.--208016, India \\
email: pallab@iitk.ac.in   \\ \mbox{} \\
and \\
J.K. Bhattacharjee\footnote{Author to whom all correspondence should be addressed}  \\
Department of Theoretical Physics, Indian Association for Cultivation of Science \\
Calcutta -- 700032, India \\
email: tpjkb@mahendra.iacs.res.in  \\ \mbox{} \\
{Dated:14th January, 2002} \\ \mbox{} \\

\end{center}
\end{large}
\baselineskip 18pt

\parindent 0.3in

  We consider photoemision from a two-electron quantum dot and find analytic expression for the
  cross-section. We show that the emission cross-section from the ground state as a function
  of the magnetic field has sharp discontinuities corresponding to the singlet-triplet transitions
  for low magnetic fields and the transitions between magic numbers (2J+1) for high magnetic field.
  We also find the corrections to the photoemission cross-section from a more relistic quantum dot having a
  finite thickness due to nonvanishing extent of electron wave function in the direction
  perpendicular to the plane in which 2-D dot electrons are usually confined.

  \vskip 0.2in

  Artificial atoms or quantum dots, which are essentially electrons confined in a two-dimensional region with
  a magnetic field in the third direction have been the subject of intense experimental and theoretical activity
  over the last few years [1]. The artificial "hydrogen atom" (single electron confined in a circular region by
  a harmonic potential with a magnetic field in the perpendicular direction) was solved for the eigenvalues and
  eigenfunctons over seventy years ago by Fock [2]. The levels were experimentally observed [3] more than fifty
  years later with the advent of quantum dots. The artificial "helium atom" (two electrons confined in a circular
  region by a harmonic potential with a magnetic field in the third direction) was tackled sixty years after the
  hydrogen atom. Maksym and Chakraborty [4] and Wagner, Merkt and Chaplik [5] worked out the energy levels and
  found an incredibly rich structure. The ground state can change its parity as the magnetic field is changed and
  there can be singlet-triplet transition. However, spectroscopic studies did not reveal the spectacular features
  because of an effect noted by these authors [4], [5]. Instead, they relied on certain thermodynamic measurements
  to support the energy level structure that they obtained. This "helium atom" problem was exactly solved a few years
  later by Dineykhan and Nazmitdinov [6], giving answers which agreed with the findings of Maksym and Cahkraborty [4]
  and Wagner et al [5]. In this note, we wish to point out that photoelectric efect is an experiment that would probe
  the different transitions in the ground state energy as the magnetic field is varied. We will show discontinuities in the cross-section as
  a function of the magnetic field corresponding to the singlet-triplet transitions and find the corrections due
  to finite thickness by imposing a stronger harmonic confinement in the z-direction.

  Considering two electrons in a circular dot with a magnetic field perpendicular to the circular region, we can write
  the hamiltonian, \begin{equation} H = \sum_{j=1}^{2} [-\frac{\hbar^{2}}{2m^{\ast}}\nabla_{j}^{2}
  +\frac{\omega_{c}}{2}(-i\hbar\nabla_{\phi_{j}})+\frac{1}{2}m^{\ast}\omega_{c}^{2}\rho_{j}^{2}]
  +\frac{e^{2}}{4\pi\epsilon\epsilon_{0}}\frac{1}{\mid \vec{\rho_{1}} -\vec{\rho_{2}}\mid} \end{equation} where
  $\vec{\rho_{1}}$ and $\vec{\rho_{2}}$ are the two dimensional position vectors of the two electrons, $\omega_{c}$ is the
  cyclotron frequency, $m^{\ast}$ is the effective mass of the electron in the semiconductor and $\Omega^{2} = (\omega_{0}^{2}
  +\frac{\omega_{c}^2}{4})$. Using the COM coordinate $\vec{\rho_{c}}=\frac{1}{2}(\vec{\rho_{1}}+\vec{\rho_{2}})$ and the relative coordinate
  $\vec{\rho_{rel}}=(\vec{\rho_{1}}-\vec{\rho_{2}})$, the hamiltonian clearly splits into $ H=H_{c}+H_{rel}$, where $H_{c}$ depends
  only on the center of mass coordinates. This part of the hamiltonian is a purely single electron hamiltonian and can be exactly
  solved. The part $H_{rel}$ on the other hand involves the important Coulomb repulsion  and is responsible for the rich structure
  of the energy spectrum. Now if there is an external electric field of long wavelength (far infrared spectroscopy) imposed on the
  system, then the dot size being much smaller than the wavelength, there is no appreciable change in the elctric field  across the
  sample and hence the contribution to the hamitonian is $e\vec{E}\cdot (\vec{\rho_{1}}+\vec{\rho_{2}})\exp(i\omega t)= 2e\vec{E}\cdot\vec{\rho_{c}}\exp(i\omega t)$
  where $\vec{E}$ is a constant electric field. Thus, this perturbation does not couple to the $\vec{\rho_{rel}}$-dependent part and
  is completely blind to the rich structure coming from. Now, if photoelectric effect is what we are interested in, then the external
  electromagnetic field can be considered as coming from a vector potential $\vec{A}^{ext}=\hat{n}A_{0}\exp(i\omega t)$ and the
  perturbation hamiltonian $\tilde{H}$ is $\frac{\vec{p}\cdot\vec{A}^{ext}}{m^{\ast}}$ to the lowest order, which can be written as
  \begin{equation}\tilde{H}=\frac{A_{0}}{m^{\ast}}[{\vec{p}}_{1}\cdot\hat{n}\exp(i\vec{k}\cdot{\vec{\rho}}_{1})+{\vec{p}}_{2}\cdot\hat{n}\exp(i\vec{k}\cdot{\vec{\rho}}_{2})]\exp(-i\omega t)\end{equation}
  and since the dipole approximation is no longer required (i.e. long wavelength condition is not imposed), we have
  the contribution to the ionization cross-section coming from both $H_{\vec{\rho_{c}}}$ and $H_{\vec{\rho_{rel}}}$.

  The photoemission cross-section involves the matrix element $\int\psi_{f}^{\ast}(\rho_{1},\rho_{2})
  \tilde{H}\psi_{i}(\rho_{1},\rho_{2})d^{2}\rho_{1}d^{2}\rho_{2}$. The initial state is the ground state of the dot and can be written
  as $\phi_{1}(\vec{\rho_{c}})\psi_{2}(\vec{\rho_{rel}})$ while the final state corresponds to a free electron of wave number $\vec{q}$ and a bound
  electron in the lowest energy state of a single electron quantum dot i.e.$\psi_{f}({\vec{\rho}}_{1},{\vec{\rho}}_{2})=\phi({\vec{\rho}}_{1})\exp(i\vec{q}\cdot {\vec{\rho}}_{2})$.
  It is this integral which is sensitive to the nature of initial state. As the initial state undergoes a singlet-triplet transition,
  the orbital parity of the spatial wave function changes and there is a sudden jump in the matrix element and hence the
  cross-section as a function of the magnetic field will show jumps at the singlet-triplet transitions.

  To establish the above result, the most important fact that we need to know is the two-electron wave function for the
  ground state. We have found an extremely accurate variational wave function for the ground state [7]. This wave function
  (normalized) is \begin{equation}\Psi(\vec{\rho_{c}},\vec{\rho_{rel}})=\frac{1}{2^{(\frac{\mid l\mid}{2}+1)}\pi\tilde{a_{H}}\beta (\Gamma(\mid l\mid +1))^{\frac{1}{2}}}
  (\frac{\rho_{rel}}{\beta})^{\mid l \mid}\exp(-\frac{\rho_{c}^{2}}{4{\tilde{a_{H}}}^{2}})\exp(-\frac{\rho_{rel}^{2}}{4{\beta}^{2}})exp(-il\phi_{rel})\end{equation}
  where ${\tilde{a_{H}}}^{2}=\frac{\hbar}{4m^{\ast}\Omega}$ and $\beta$ is a variational parameter fixed by the energy minimization
  condition \begin{equation}x^{4}-\sqrt{2} \, \frac{\tilde{a_{H}}}{a^{\ast}}\frac{\Gamma(\mid l \mid +\frac{1}{2})}{\Gamma(\mid l \mid+2)}x-1=0\end{equation}
  where $x=\frac{\beta}{2\tilde{a_{H}}}$ and $a^{\ast}=\frac{4\pi\epsilon\epsilon_{0}\hbar^{2}}{m^{\ast}e^{2}}$ is the Bohr radius
  of the semicoductor material (energy spectrum is plotted in fig.1). The final state corresponds to one free electron, and one electron in the ground state of single electron quantum dot.

  The matrix element for photoemission can now be written as\begin{equation}<f|\tilde{H}|i>
  =\frac{e}{m^{\ast}}\sqrt{\frac{2\pi\hbar}{\omega}}[\int\!\!\int\psi^{\star}_{final}({\vec{\rho}}_{1},{\vec{\rho}}_{2})
  \exp(i\vec{k}\cdot{\vec{\rho}}_{1})(-i\hbar\hat{n}\cdot\nabla_{1})\psi_{initial}({\vec{\rho}}_{1},{\vec{\rho}}_{2})\,d{\vec{\rho}}_{1}\,d{\vec{\rho}}_{2}
  +(1\longleftrightarrow2)]\end{equation} The two contributions to $<f|\tilde{H}|i>$ will be equal and hence we need to
  evaluate only one integral I which is
  \begin{eqnarray}I & = & \frac{e}{m^{\ast}}\sqrt{\frac{2\pi\hbar}{\omega}}\frac{(i)^{2}\hbar\hat{n}\cdot(\vec{k}-\vec{q})}{\sqrt{2^{(\mid l \mid+3)}\pi^{3}a_{H}^{2}{\tilde{a_{H}}^{2}\beta^2\Gamma(\mid l\mid+1)A}}}\int\!\!\int\exp[i(\vec{k}-\vec{q})\cdot(\vec{\rho_{1}})]\exp[-\frac{\rho_{2}^{2}}{4a_{H}^{2}}]\exp[-\frac{\rho_{c}^{2}}{4{\tilde{a_{H}}}^{2}}]\times \nonumber \\
                    &   & \exp[-\frac{\rho_{rel}^{2}}{4\beta^{2}}-il\phi_{rel}](\frac{\rho_{rel}}{\beta})^{\mid l \mid}d^{2}\rho_{1}d^{2}\rho_{2}\end{eqnarray}
  where $a_{H}^{2}=2{\tilde{a_{H}}}^{2}$.
  Now, \begin{eqnarray}\exp[-\frac{\rho_{c}^{2}}{4{\tilde{a_{H}}}^{2}}]\exp[-\frac{\rho_{rel}^{2}}{4\beta^{2}}-il\phi_{rel}](\frac{\rho_{rel}}{\beta})^{\mid l \mid} & = & \exp[-(\rho_{1}^{2}+\rho_{2}^{2})(\frac{1}{16{\tilde{a_{H}}}^{2}}+\frac{1}{4\beta^{2}})]\exp[-\rho_{1}\rho_{2}(\frac{1}{8{\tilde{a_{H}}}^{2}}-\frac{1}{2\beta^{2}})\times \nonumber \\
                                                                                                                                                                     &   & \cos(\phi_{1}-\phi_{2})](\frac{\rho_{1}\exp[-i\phi_{1}]-\rho_{2}\exp[-i\phi_{2}]}{\beta})^{(\mid l \mid)}\end{eqnarray} After putting the expression in equation (7) in equation (6) and
 writing $\vec{k}-\vec{q}= \vec{K}$, we note that  $\vec{K}\cdot\vec{\rho_{1}}=K\rho_{1}\cos(\phi_{1}-\eta)$. Now the angular integrations in the above integral can be performed
 by using, as necessary,the following identities: \begin{eqnarray}\exp[ib\cos(\phi_{1}-\phi_{2})] & = & \sum_{m=-\infty}^{\infty}(i)^{m}J_{m}(b)\exp[im(\phi_{1}-\phi_{2})] \nonumber \\
                                                                 \exp[-c\cos(\phi_{1}-\phi_{2})] & = & \sum_{m=-\infty}^{\infty}(i)^{2m}I_{m}(c)\exp[im(\phi_{1}-\phi_{2})] \nonumber \\
                                                                 \int_{0}^{2\pi}\frac{d\phi}{2\pi}\exp[i(m-n)\phi] & = & \delta_{m,n} \end{eqnarray}
After carrying out the angular integrals with the help of binomial expansion
\begin{equation} (\frac{\rho_{1}\exp[-i\phi_{1}]-\rho_{2}\exp[-i\phi_{2}}{\beta})^{\mid l \mid}  =  \sum_{t=0}^{\mid l \mid}\frac{\Gamma(\mid l \mid+1)}{\Gamma(t+1)\Gamma(\mid l \mid -t+1)}\frac{\rho_{1}^{\mid l \mid -t}\rho_{2}^{t}}{\beta^{\mid l \mid}}\exp[-i(\mid l \mid-t)\phi_{1}]\exp[-it\phi_{2}] \end{equation}
and writing the constant  term outside as $C_{l}$ we are left with radial integral parts \begin{eqnarray}<f\mid \tilde{H} \mid i> & = &  2\frac{C_{l}}{\beta^{\mid l \mid}}\int\rho_{1}d\rho_{1}\rho_{2}d\rho_{2}\sum_{t=0}^{l} C_{t}^{\mid l \mid} \rho_{1}^{\mid l \mid -t}\rho_{2}^{t}\exp[-\rho_{1}^{2}b]\exp[-\rho_{2}^2c](-1)^{t}(i)^{\mid l \mid} \times \nonumber \\
                                                                              &   & \exp[-i\mid l \mid \eta (2 \pi)^{2}J_{l}(K\rho_{1})I_{t}(d\rho_{1}\rho_{2})] \\
                                                                              & = & \frac{(2\pi)^{2}C_{l}}{4\beta^{\mid l \mid} bc}\sum_{t=0}^{l}(-1)^t C_{t}^{l}(i)^{l}\exp[-il\eta]\frac{1}{(\sqrt{b})^{l-t}\sqrt{c}^{t}}\sum_{p=0}^{\infty}\sum_{q=0}^{\infty}(-1)^{p}\frac{\Gamma(p+q+ \mid l \mid+1)}{\Gamma(p+1)\Gamma(q+1)\Gamma(\mid l \mid +p+1)}\times \nonumber \\
                                                                              &   & (\frac{K}{2\sqrt{b}})^{2p+\mid l \mid}(\frac{d}{2 \sqrt{bc}})^{2q+t} \end{eqnarray}
After some manipulations and using the generating function of associated Laguerre polynomial we find the closed form
\begin{equation}<f\mid \tilde{H} \mid i> = \frac{(2\pi)^{2}C_{l}}{4\beta^{\mid l \mid} bc}(\frac{K}{2b})^{\mid l \mid}\exp[-il\eta+i\frac{l\pi}{2}]\frac{(1-\frac{d}{2c})^{\mid l \mid}}{(1-\frac{d^{2}}{4bc})^{\mid l \mid +1}}\exp[-\frac{K^{2}}{4b(1-\frac{d^{2}}{4bc})}]\end{equation}
where $b=\frac{1+\frac{2}{x^{2}}}{8a_{H}^{2}}$, $c=\frac{3+\frac{2}{x^{2}}}{8a_{H}^{2}}$ and $d=\frac{1-\frac{2}{x^{2}}}{4a_{H}^{2}}$. From this we get the expression for the differential cross-section
\begin{equation}\frac{\frac{d\sigma}{d\phi_{\vec{q}}}}{\sigma_{0}} = \frac{2^{8+3\mid l \mid}}{\Gamma(\mid l \mid +1)}[1-(\mid l \mid +1)\frac{\Omega}{\omega}+\frac{l}{2}\frac{\omega_{c}}{\omega}]
\sin^{2}\theta\cos^{2}\phi\frac{(x^{2}+2)^{2\mid l \mid}(Ka_{H})^{2\mid l \mid}}{x^{2\mid l \mid-2}(x^{2}+6)^{2\mid l \mid+2}}\exp[-2\frac{K^{2}a_{H}^{2}(3x^{2}+2)}{(6+x^{2})}]\end{equation}
Here $\vec{q}$ is the wave vector of the emitted electron, $\vec{k}$ is the wave vector of the incident
photon, $\hbar\vec{K}=\hbar(\vec{k}-\vec{q})$ is the momentum transferred and
$\theta$ and $\phi$ are respectively the angles $\vec{q}$ makes with $\vec{k}$
and $\vec{k}\hat{n}$ plane where $\hat{n}$ is the unit polarization vector of
the incident photon. $\sigma_{0}=\frac{e^{2}}{c{a^{\ast}}^{2}}(\frac{m_{e}}{m^{\ast}})^{2}(\frac{2\pi}{\hbar})$ is a
constant extracted to express the differntial cross-section expression in a dimensionless form.
So, \begin{equation}K^{2}=k^{2}+q^{2}-2kq\cos\theta \end{equation} holds. and $\cos\phi=\sin\theta_{\hat{n}}\cos(\phi_{\hat{n}}-\phi_{\vec{q}})$
and $\cos\theta=\sin\theta_{\vec{k}}\cos(\phi_{\vec{k}}-\phi_{\vec{q}})$. Therefore, it is very clear
from the expression for the differential cross-section that it depends significantly
on the direction of incidence and polarization. This for some simple cases can be
illustrated easily. If $\vec{k}$ is parallel to z-axis then $\cos\theta=0$ and
$\cos\phi=\cos\phi_{\vec{q}}$ or $\cos\phi=\sin\phi_{\vec{q}}$ as $\hat{n}$ is parallel to x or
y-axis. So, the angular distribution is proportional to $\cos^{2}\phi_{\vec{q}}$ or $\sin^{2}\phi_{\vec{q}}$ and if
it is the case of circular polarization then the angular distribution is proportional to $(\cos^{2}\phi_{\hat{n}}\cos^{2}\phi_{\vec{q}}+\sin^{2}\phi_{\hat{n}}\sin^{2}\phi_{\vec{q}})$ and
only if $\hat{n}=\frac{(\hat{x}\pm i\hat{y})}{\sqrt{2}}$ then it becomes isotropic.
But, when the $\vec{k}$ lies in the x-y plane then with all the cases of circular polarization we shall
have angular dependence. It also becomes apparent from the expression that emission count is
larger in the direction of polarization compared to other cases and if the photon is
linearly polarized in the z-direction then there is no emission. So, depending on
the 'l' values of the ground state as a function of magnetic field the cross-section
would have different angular distribution as well as discontinuities characterizing
transitions of the ground state. In this case it will also be found that if Based on these one can probe now these transitions
experimentally. For this purpose one has to choose carefully the magnetic field strength, incident photon frequency and the abovementioned directions. We plot
the dimensionless expression of the differential cross-section (fig.2) for the transitions $l=0\rightarrow l=1$ and $l=1\rightarrow l=2$ which take place (for our
chosen system size ${\hbar\omega_{0}=4meV}$) at 1.3T and 6.1T respectively when $\vec{k}\parallel \hat{z}$ and $\vec{q} \parallel \hat{n}$ (i.e. when it is maximum).
From the plot it can be seen that there are discontinuities at the magnetic field strengths where singlet-triplet transitions are taking place and where
for the cases of $l=0$ and $l=1$ the cross-section increases with magnetic field strength for $l=2$ it decreases. This can be understood on the physical ground
in analogy with atomic photo-effect. As magnetic field strength increases wave function is compressed more and more that means electrons
become more tightly bound and emission increases. When transition takes place depending on the energetics electron wave function becomes
further compressed which manifests in the x-values for different 'l' values and the emission count shows a jump. But, after sufficient increase in the magnetic field strength
the energy value also changes considerably and with it ionization energy changes. Thus upon keeping the incident frequency
fixed after a certain range of magnetic field the cross-section decreases with increasing field strength.

 In the real dots there is finite extent of the wave function in the z-direction and for the experiments it also needs to be considered. We teke this into account
 by considering a stronger harmonic confinement in the z-direction. with this the hamiltonian gets modified with the term
 \begin{equation}
H_{z} = -\frac{\hbar^{2}}{2m^{\ast}}\frac{d^{2}}{dz^{2}}+\frac{1}{2}m^{\ast}\omega_{z}^{2}z^{2}
\end{equation}
and the Coulomb term becomes
$\frac{e^{2}}{4\pi\epsilon\epsilon_{0}}\frac{1}{\mid \vec{r_{1}} -\vec{r_{2}}\mid}$.
With the $H_{z}$ part we have the wavefunction modified by
\begin{equation}\psi_{3}(z)=\frac{1}{\sqrt{2^{n}\Gamma(n_{z}+1)\pi^{1/2}\lambda}}\exp(-\frac{z^{2}}{2\lambda^{2}})H_{n}(\frac{z}{\lambda})\end{equation}
and the energy is modified by the term $E_{z}=(n_{z}+\frac{1}{2})\hbar\omega_{z}$. When solved variationally (introducing two parameters $\beta_{1}$ and $\beta_{2}$ instead of two independent oscillator lengths) we have for the two-electron dot the following
coupled characteristic equations
\begin{equation}x^{4}[1-\frac{2}{y^{3}}(\frac{a_{H}}{a^{\ast}})(\frac{a_{H}}{\lambda_{rel}})^{3}\frac{\Gamma(\mid l \mid+1)}{\Gamma(\frac{5}{2}+\mid l \mid)}2F_{1}(\frac{3}{2},2+\mid l \mid,\frac{5}{2}+\mid l \mid,1-2(\frac{x}{y})^{2}(\frac{a_{H}}{\lambda_{rel}})^{2})]=1\end{equation}
\begin{eqnarray}y^{4}-y(\frac{\lambda_{rel}}{a^{\ast}})\frac{\Gamma(\mid l \mid+1)}{\Gamma(\frac{3}{2}+\mid l \mid)}2F_{1}(\frac{1}{2},1+\mid l \mid,\frac{3}{2}+\mid l \mid,1-2(\frac{x}{y})^{2}(\frac{a_{H}}{\lambda_{rel}})^{2})+2\frac{x^{2}}{y}\frac{a_{H}^{2}}{a^{\ast}\lambda_{rel}}\frac{\Gamma(\mid l \mid+2)}{\Gamma(\frac{5}{2}+\mid l \mid)}\times \nonumber \\
2F_{1}(\frac{3}{2},2+\mid l \mid,\frac{5}{2}+\mid l \mid,1-2(\frac{x}{y})^{2}(\frac{a_{H}}{\lambda_{rel}})^{2})=1\end{eqnarray} where oscilator lengths are related as $\lambda^{2}=\frac{\lambda_{rel}^{2}}{2}=\frac{\hbar}{m^{\ast}\omega_{z}}$ and $y=\frac{\beta_{2}}{\lambda_{rel}}$. When solved numerically, from the energy spectrum it is found that transitions of the ground state are taking place at
higher magnetic field values ($l=0\rightarrow l=1$ and $l=1\rightarrow l=2$ occur at 3.1T and 11.2T respectively for our chosen ratio $\frac{\omega_{z}}{\omega_{0}}=9$) as it becomes evident from fig.5 and the dimensionless differential cross-section now becomes
\begin{eqnarray}\frac{\frac{d\sigma}{d\Omega_{\vec{q}}}}{\sigma_{0}} & = & \frac{2^{8+3\mid l \mid}}{\Gamma(\mid l \mid +1)}[1-(\mid l \mid +1)\frac{\Omega}{\omega}+\frac{l}{2}\frac{\omega_{c}}{\omega}]\sin^{2}\theta\cos^{2}\phi\frac{(x^{2}+2)^{2\mid l \mid}(K_{\rho}a_{H})^{2\mid l \mid}}{x^{2\mid l \mid-2}(x^{2}+6)^{2\mid l \mid+2}}\exp[-2\frac{K_{\rho}^{2}a_{H}^{2}(3x^{2}+2)}{(6+x^{2})}]\times \nonumber \\
                                                                     &   & \frac{8}{\sqrt{2\pi}}\frac{1}{y}(\frac{\lambda_{rel}}{a^{\ast}})\frac{1}{(\frac{3}{y^{4}}+\frac{20}{y^2}+8)}\exp[-K_{z}^{2}\lambda_{rel}^{2}(\frac{1}{(3+\frac{1}{y^{2}})}+\frac{(2-\frac{1}{y^{2}})^{2}}{4(3+\frac{1}{y^{2}})(1+\frac{1}{y^{2}})-(2-\frac{1}{y^{2}})^{2}})]\end{eqnarray} Here, $K_{\rho}^{2}=K_{x}^{2}+K_{y}^{2}$ and $K_{z}^{2}=K^{2}-K_{\rho}^{2}=K^{2}\cos^{2}\theta_{\vec{K}}$ and
\begin{eqnarray}\cos\theta_{\vec{K}} & = & \frac{k\cos\theta_{\vec{k}}-q\cos\theta_{\vec{q}}}{\sqrt{k^{2}+q^{2}-2kq\cos\theta}} \nonumber \\
                \cos\theta & = & \sin\theta_{\vec{k}}\sin\theta_{\vec{q}}\cos(\phi_{\vec{k}}-\phi_{\vec{q}})+\cos\theta_{\vec{k}}\cos\theta_{\vec{q}} \nonumber \\
                \sin\theta\cos\phi & = & \sin\theta_{\hat{n}}\sin\theta_{\vec{q}}\cos(\phi_{\hat{n}}-\phi_{\vec{q}})+\cos\theta_{\hat{n}}\cos\theta_{\vec{q}} \end{eqnarray} Putting these in the expression for differential cross-section and by explicit integration one can find the total cross-section. For the incidence direction parallel to z-axis and emission in the
direction of polarization we show the plot for this modified differential cross-section (fig.7) with the earlier photon energy. From the plot it is seen that the cross-section value is now sufficiently suppressed compared to earlier case as energy of the corresponding states have significantly. Also, due to this reason where in the earlier 2-D case decrease in the cross-section as a function of field strength
started for $l=2$ for the 3-D dot it started decreasing right from the $l=1$ state after certain amount of increase and at $l=1\rightarrow l=2$ transition the count decreased in contrast to increase in 2-D situation.

We conclude by considering the feasibility of an experiment which would detect the above effect. The first thing to note is that we want a coupling of relative co-ordinate
to the external electromagnetic field.Consequently the wavelength of the radiation must be smaller than the size of the dot which can be of the order of 100 nanometers.
This implies that we will be dealing with energetic photons and photo-emission will be from the bound state  in the dot into vacuum. There will be emission from the semiconductor
as well and we would have to substract this background contribution.This can be achieved by studying the angular distribution.The distribution of the electron knocked out from the
rest of the solid will be isotropic where as the distribution of the electrons coming from the dot would have a zero in the forward direction. This enables one to know the background
which can now be substracted to find the cross-section of the photoemission from the dot. Another point is that as the magnetic field applied to the dot can be considered to be local
compared to the bulk, with the change of magnetic field while there will be change in the angular dependence and counts from the dot, that from the bulk would remain unchanged.

An alternative technique is to do the experiment first with the GaAs layer without dot. This determine the background. Next,one can
repeat it with single-electron dots and finally with two-electron dots. The difference between the two-electron dot and the single-electron dot
will exhibit the correlation effect.

From the plot for angular distribution for differential cross-section (fig.3,fig.4) it becomes clear that when transitions occur the angular distribution also changes and maximum of emission is obtained for a definite angle of incidence and this angle changes with transitions. So, at the time of experiments keeping the photon enegy fixed and
varying the angle of incidence and observing the change in angle corresponding to maxima due to change in 'l' states, the transitions can be probed. The most significant point which should be stressed is that for certain angle of incidence (as evident from fig.3 and fig.4) the count does not change even at the points of transitions. So to detect the transitions by the discontinuities in the cross-section, this angle should be carefully avoided. We have already argued above that
if photon energy is kept fixed then all the transitions can not be observed properly. So, depending on the system size and field strength and 'l' values of the states angle of incidence and photon energy have to be carefully chosen and maximum of count will always be obtained in the direction of polarization.

Is the effect big enough to be measured? The scale for the cross-section of emission from the dot when compared to the scale for the
emission from the bound state of an atom is the ratio $\frac{m}{m^{\ast}}\frac{\tilde{a_{H}}^{2}}{\beta^{2}}$ which is much greater than unity and from our analytic expression
for the emission cross-section it is evident that the ratio $ \sigma_{0}$ is a large number. So the scale for the  emission cross-section is bigger when emission
from the dot is involved. This should make the experiment quite feasible.

\bibliographystyle{plain}

\begin{figure}
\begin{center}
\epsfig{file=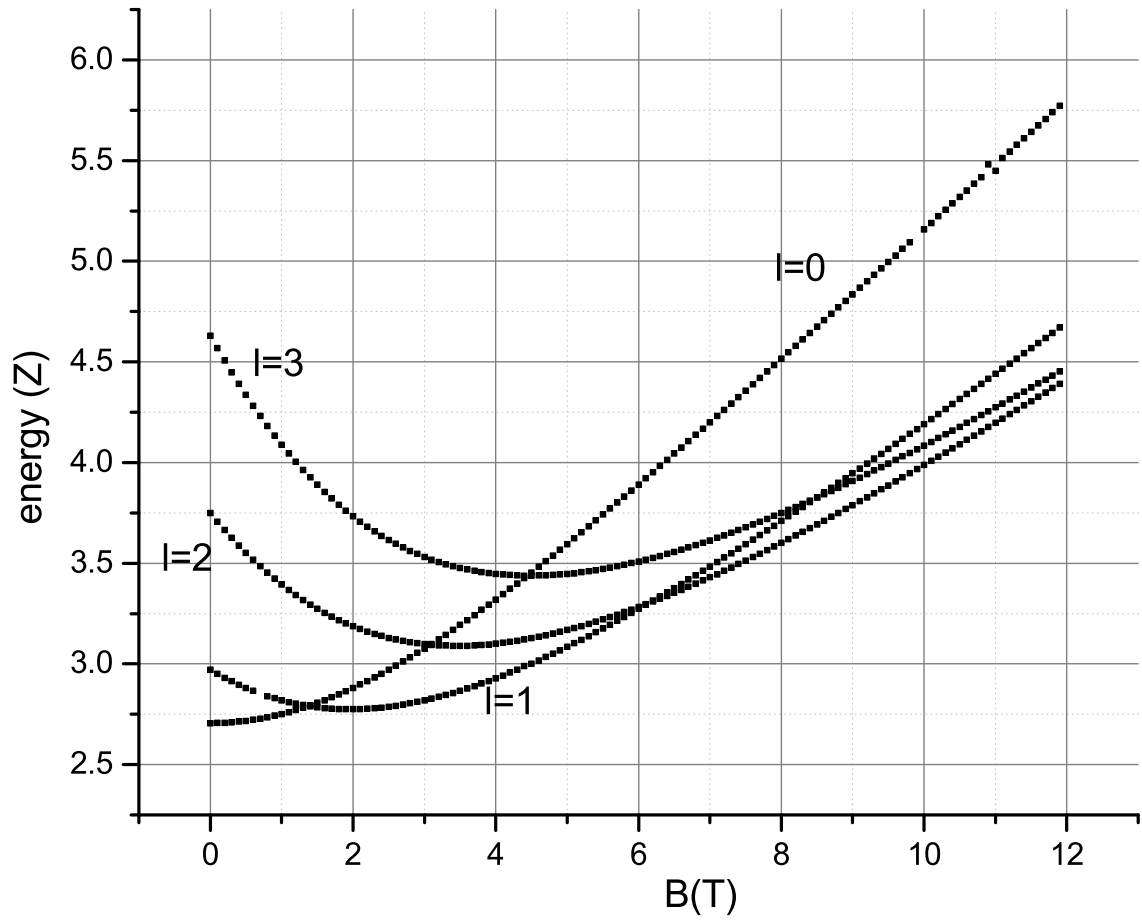,width=1.0\linewidth}
\end{center}
\vskip 0.5in
\caption{$Z=\frac{E_{rel}+E_{spin}}{\hbar\omega_{0}}$ vs. B(T) is plotted for different 'l' values and $l=0\rightarrow l=1$, $l=1\rightarrow l=2$ transitions are taking place at $B=1.3T$ and $B=6.1T$ respectively and also other energy level crossings are present.}
\label{fig:fig1}
\end{figure}

\begin{figure}
\begin{center}
\epsfig{file=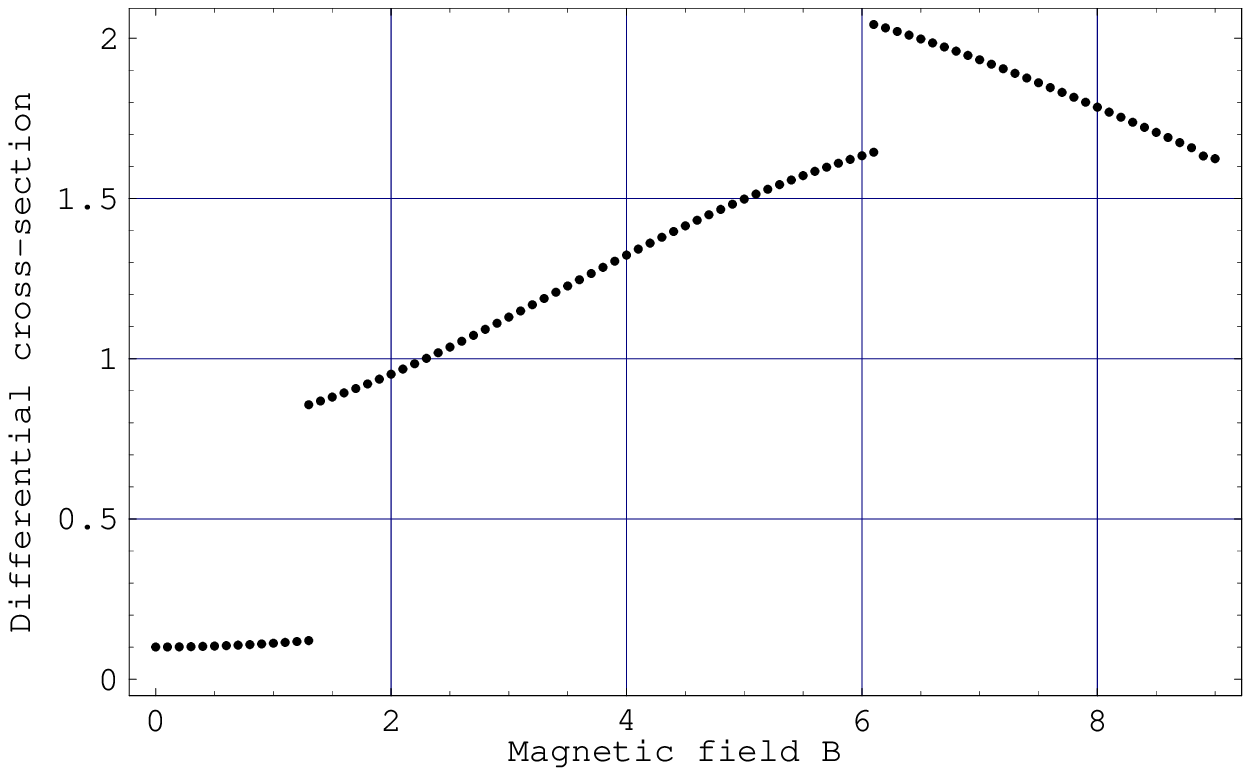,width=1.0\linewidth}
\end{center}
\vskip 0.5in
\caption{$\frac{\frac{d\sigma}{d\phi_{\vec{q}}}}{\sigma_{0}}$ vs. B(T) is plotted for different 'l' values of the ground state as the B is varied and showing discontinuities as characteristic of the transitions. From the plot it is found that percentage change for $l=1 \rightarrow l=2$ is $\approx 4.5$ times smaller compared to $l=0 \rightarrow l=1$ transition and also it is evident from the plot that at a fixed frequency behaviors of different 'l' cross-sections are going to change and for this reason both $\omega$ and B have to be varied to observe all the transitions properly.}
\label{fig:fig2}
\end{figure}

\begin{figure}
\begin{center}
\epsfig{file=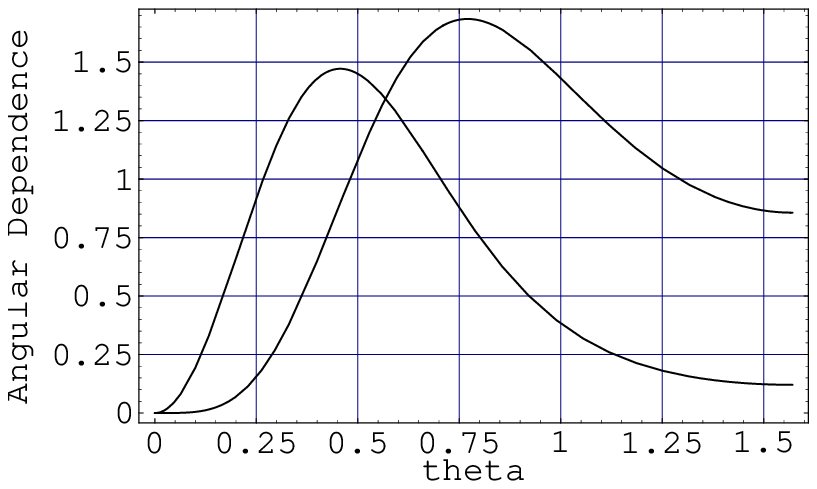,width=1.0\linewidth}
\end{center}
\vskip 0.5in
\caption{Difference in angular dependence of $\frac{\frac{d\sigma}{d\phi_{\vec{q}}}}{\sigma_{0}}$ for transition $l=0\rightarrow l=1$ is shown}
\label{fig:fig3}
\end{figure}

\begin{figure}
\begin{center}
\epsfig{file= 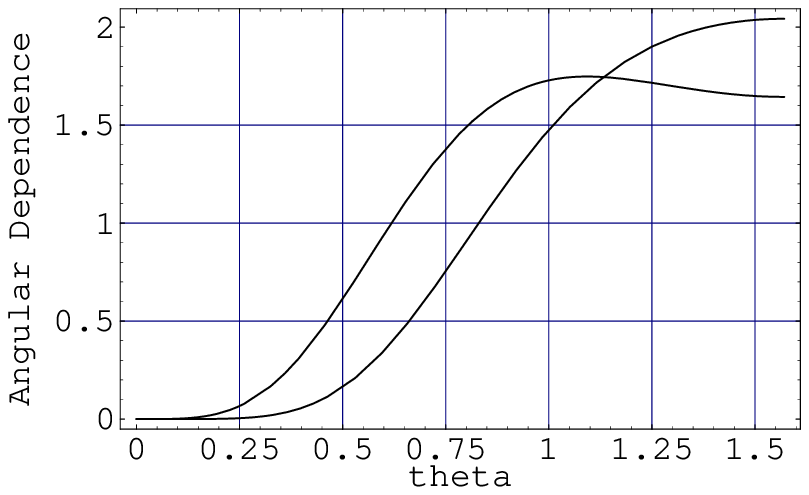,width=1.0\linewidth}
\end{center}
\vskip 0.5in
\caption{Difference in angular dependence of $\frac{\frac{d\sigma}{d\phi_{\vec{q}}}}{\sigma_{0}}$ for transition $l=1\rightarrow l=2$ is shown}
\label{fig:fig4}
\end{figure}

\begin{figure}
\begin{center}
\epsfig{file=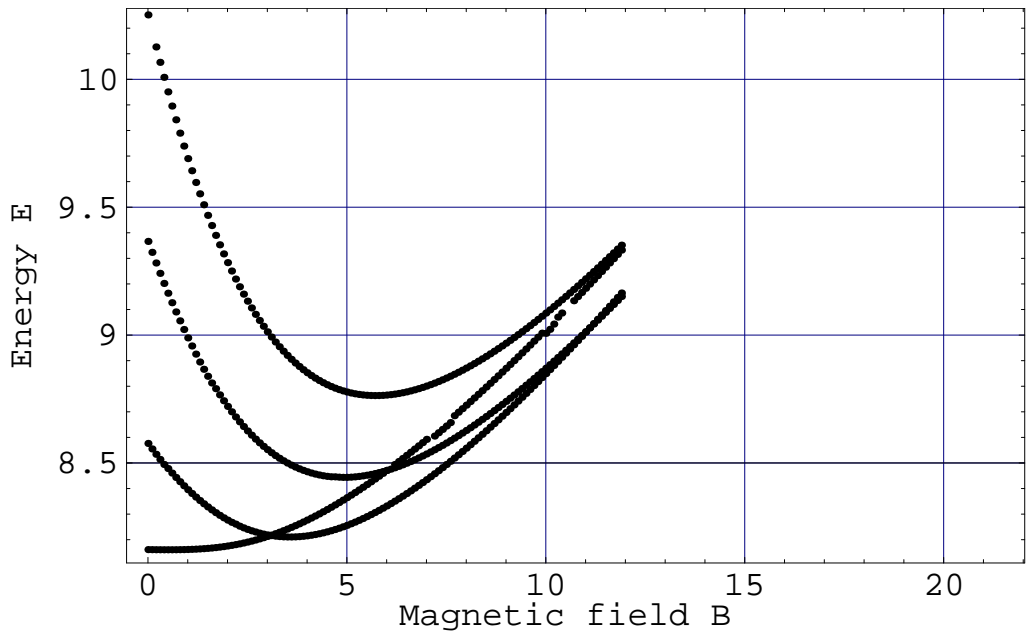,width=1.0\linewidth}
\end{center}
\vskip 0.5in
\caption{$E=\frac{E_{rel}+E_{spin}}{\hbar\omega_{0}}$ vs. B(T) plotted for different 'l' values when finite thickness of the dot is taken into account.For this purpose $\frac{\omega_{z}}{\omega_{0}}=9$ has been taken.}
\label{fig:fig5}
\end{figure}

\begin{figure}
\begin{center}
\epsfig{file=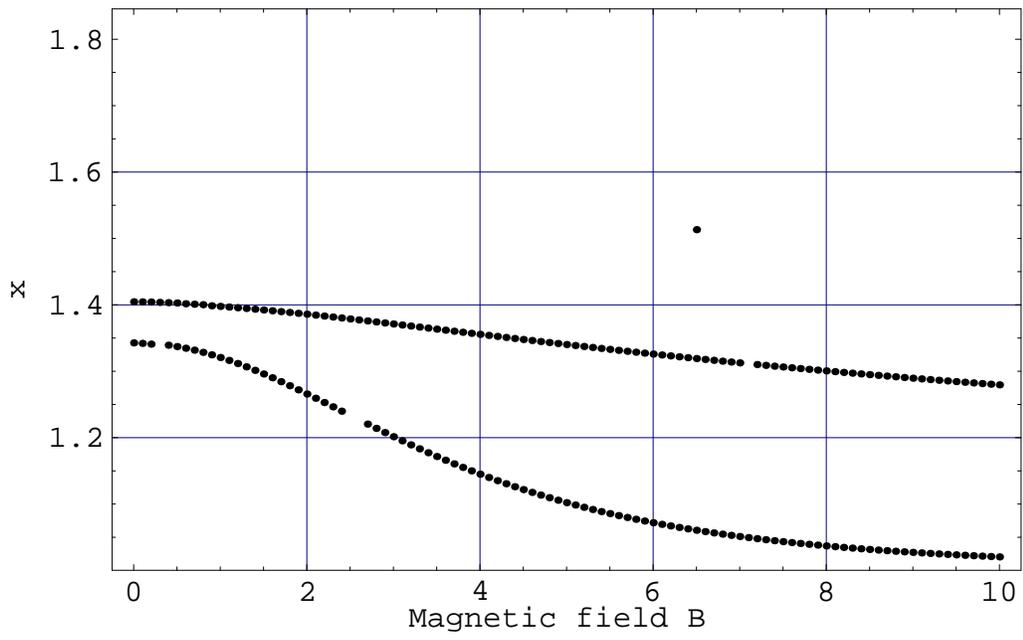,width=1.0\linewidth}
\end{center}
\vskip 0.5in
\caption{$x$ vs. B(T) is plotted for $l=0$ for both the 2-D and realistic 3-D dot and it is clearly found that throughout the range of magnetic field x values have decreased for the later one. }
\label{fig:fig6}
\end{figure}

\begin{figure}
\begin{center}
\epsfig{file=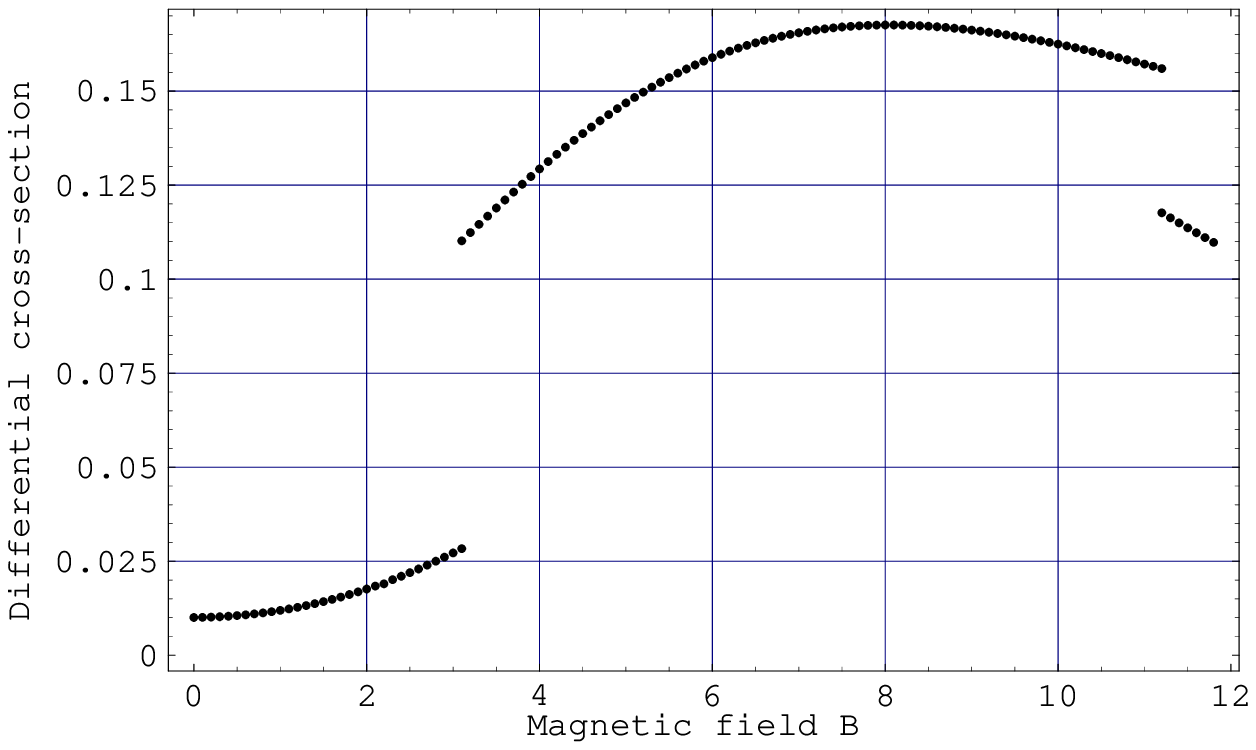,width=1.0\linewidth}
\end{center}
\vskip 0.5in
\caption{$\frac{\frac{d\sigma}{d\Omega_{\vec{q}}}}{\sigma_{0}}$ vs. B(T) is plotted for different 'l' values of the ground state as the B is varied and showing discontinuities as characteristic of the transitions. From the plot it is found that percentage change for $l=1 \rightarrow l=2$ is $\approx 4.5$ times smaller compared to $l=0 \rightarrow l=1$ transition and also it is evident from the plot that at a fixed frequency behaviors of different 'l' cross-sections are going to change and for this reason both $\omega$ and B have to be varied to observe all the transitions properly.}
\label{fig:fig7}
\end{figure}

\end{document}